\newif\ifpdflatex    % toggle between pdf and ps images
\pdflatextrue           % use with pdflatex (plat.sh)
\documentclass[fleqn,usenatbib,useAMS,twocolumn,numberedappendix]{aastex}
\usepackage{amsmath,esint}
\usepackage{url,aasmacros}
\usepackage{natbib}
\usepackage{soul, CJK}
\usepackage{multirow}
\usepackage{tablefootnote}
\usepackage{appendix}
\usepackage{bm}
\usepackage{xspace}
\usepackage{color}
\usepackage{enumitem}
\usepackage{booktabs}

\usepackage{graphicx}	% Including figure files
\usepackage{amsmath}	% Advanced maths commands
\usepackage{amssymb}	% Extra maths symbols
\usepackage{bm}		% Bold maths symbols, including upright Greek
\usepackage{url}
\usepackage{amsbsy}
\usepackage{xspace}
\usepackage{hyperref}
\usepackage{longtable}
\usepackage[colorinlistoftodos]{todonotes}
\usepackage[flushleft]{threeparttable}

\usepackage[T1]{fontenc}
\usepackage{ae,aecompl}
\def\lesssim{\mathrel{\hbox{\rlap{\hbox{\lower5pt\hbox{$\sim$}}}\hbox{$<$}}}}
\def\gtrsim{\mathrel{\hbox{\rlap{\hbox{\lower5pt\hbox{$\sim$}}}\hbox{$>$}}}}

\usepackage{upgreek}                                                  
\usepackage{xspace}                                                       
\def\apjl{ApJ}%
\def\apjs{ApJS}%
\def\aap{A\&A}%
\shorttitle{WINTER on S250206dm}

\begin{document}
\title{WINTER on S250206dm: A near-infrared search for an electromagnetic counterpart to a gravitational-wave event}

\author[0000-0002-7197-9004]{Danielle Frostig}
\altaffiliation{These authors contributed equally to this work.}
\affil{Center for Astrophysics | Harvard \& Smithsonian, 60 Garden Street, Cambridge, MA 02138, USA}

\author[0000-0003-2758-159X]{Viraj R. Karambelkar}
\altaffiliation{These authors contributed equally to this work.}
\affil{Cahill Center for Astrophysics, California Institute of Technology, Pasadena, CA 91125, USA}

\author[0000-0003-2434-0387]{Robert D. Stein}
\altaffiliation{These authors contributed equally to this work.}
\affil{Department of Astronomy, University of Maryland, College Park, MD 20742, USA}
\affil{Joint Space-Science Institute, University of Maryland, College Park, MD 20742, USA} 
\affil{Astrophysics Science Division, NASA Goddard Space Flight Center, Mail Code 661, Greenbelt, MD 20771, USA}
%%\thanks{These authors contributed equally to this work.}
%\altaffiliation{These authors contributed equally to this work.}

\author[0000-0002-4585-9981]{Nathan~P.~Lourie}
\affiliation{MIT-Kavli Institute for Astrophysics and Space Research, Massachusetts Institute of Technology, 77 Massachusetts Ave, Cambridge, MA 02139, USA} 

\author[0000-0002-5619-4938]{Mansi M. Kasliwal}
\affil{Cahill Center for Astrophysics, California Institute of Technology, Pasadena, CA 91125, USA}

\author[0000-0003-3769-9559]{Robert A. Simcoe}
\affiliation{Department of Physics, Massachusetts Institute of Technology, 77 Massachusetts
Ave, Cambridge, MA 02139, USA} 
\affiliation{MIT-Kavli Institute for Astrophysics and Space Research, Massachusetts Institute of Technology, 77 Massachusetts Ave, Cambridge, MA 02139, USA} 

\author[0000-0002-8255-5127]{Mattia Bulla}
\affil{Department of Physics and Earth Science, University of Ferrara, via Saragat 1, I-44122 Ferrara, Italy}
\affil{INFN, Sezione di Ferrara, via Saragat 1, I-44122 Ferrara, Italy}
\affil{INAF, Osservatorio Astronomico d’Abruzzo, via Mentore Maggini snc, 64100 Teramo, Italy}

\author[0000-0002-2184-6430]{Tom\'{a}s Ahumada}
\affil{Cahill Center for Astrophysics, California Institute of Technology, Pasadena, CA 91125, USA}

\author[0000-0001-6331-112X]{Geoffrey~Mo}
\affiliation{Department of Physics, Massachusetts Institute of Technology, 77 Massachusetts
Ave, Cambridge, MA 02139, USA} 
\affiliation{MIT-Kavli Institute for Astrophysics and Space Research, Massachusetts Institute of Technology, 77 Massachusetts Ave, Cambridge, MA 02139, USA} 
\affiliation{MIT LIGO Laboratory, Massachusetts Institute of Technology, Cambridge, MA 02139, USA}

\author[0000-0003-1227-3738]{Josiah Purdum}
\affil{Cahill Center for Astrophysics, California Institute of Technology, Pasadena, CA 91125, USA}

\author[0000-0001-5926-3911]{Jill Juneau}
\affiliation{MIT-Kavli Institute for Astrophysics and Space Research, Massachusetts Institute of Technology, 77 Massachusetts Ave, Cambridge, MA 02139, USA} 

\author{Andrew Malonis}
\affiliation{MIT-Kavli Institute for Astrophysics and Space Research, Massachusetts Institute of Technology, 77 Massachusetts Ave, Cambridge, MA 02139, USA} 

\author[0000-0001-8467-9767]{G\'abor F\H{u}r\'esz}
\affiliation{MIT-Kavli Institute for Astrophysics and Space Research, Massachusetts Institute of Technology, 77 Massachusetts Ave, Cambridge, MA 02139, USA}

\begin{abstract}
%We present near-infrared follow-up observations of the International Gravitational Wave Network (IGWN) event S250206dm with the Wide-Field Infrared Transient Explorer (WINTER). WINTER is a near-infrared time-domain survey designed for electromagnetic follow-up of gravitational-wave sources localized to $\leq300\,\text{deg}^{2}$. The instrument's wide field of view (1.2 deg$^2$), dedicated 1-m robotic telescope, and near-infrared coverage (0.9-1.7 microns) are optimized for searching for kilonovae, which are expected to exhibit a relatively long-lived near-infrared component. S250206dm is the only neutron star merger in the fourth observing run (to date) localized to $\leq300\,\text{deg}^{2}$ with a False Alarm Rate below one per year. It has a $55\%$  probability of being a neutron star–black hole (NSBH) merger and a $37\%$  probability of being a binary neutron star (BNS) merger, with a $50\%$  credible region spanning 38\,deg$^2$, an estimated distance of 373\,Mpc, and an overall false alarm rate of approximately one in 25 years. WINTER covered $43\%$ of the probability area at least once and $35\%$ at least three times. Through automated and human candidate vetting, all transient candidates found in WINTER coverage were rejected as kilonova candidates. Unsurprisingly, given the large estimated distance of 373\,Mpc, the WINTER upper limits do not constrain kilonova models. This study highlights the promise of systematic infrared searches and the need for future wider and deeper infrared surveys. \\ 

We present near-infrared follow-up observations of the International Gravitational Wave Network (IGWN) event S250206dm with the Wide-Field Infrared Transient Explorer (WINTER). Near-infrared observations are a critical component of electromagnetic follow-up to gravitational-wave events, as kilonovae are expected to exhibit long-lived emission at these wavelengths, especially from lanthanide-rich ejecta. WINTER is a near-infrared time-domain survey facility designed for EM follow-up of gravitational-wave sources, featuring a wide field of view (1.2 deg$^2$), a dedicated 1-m robotic telescope, and coverage spanning 0.9–1.7 microns. S250206dm is the only neutron star merger in the fourth observing run, to date, localized to $\leq300\,\text{deg}^{2}$ with a False Alarm Rate below one per year, making it a particularly valuable target for follow-up. It has a 55\% probability of being a neutron star–black hole (NSBH) merger and a 37\% probability of being a binary neutron star (BNS) merger. The event’s estimated distance is 373 Mpc, with a 50\% credible region spanning 38 deg$^2$. WINTER covered 43\% of the probability area at least once and 35\% at least three times. Through automated and human candidate vetting, all transients were rejected as kilonova candidates. Given the large distance of the event, the WINTER upper limits do not place meaningful constraints on kilonova models. However, similar observations of future events—or in combination with optical surveys—can begin to exclude portions of the kilonova model space. This study highlights the promise of systematic infrared searches and the need for future wider and deeper infrared surveys. \\
\end{abstract}

\section{Introduction}
%\textcolor{blue}{Usual, (but not too long) schpiel about GW events, previous works, etc. Benefits of infrared, lack of observatories. Description of WINTER and this trigger.  }

%The discovery of GW170817 during the second Advanced LIGO-Virgo observing run (O2) marked a breakthrough in multi-messenger astronomy, providing the first confirmed detection of a kilonova powered by r-process nucleosynthesis in the neutron-rich ejecta of a binary neutron star (BNS) merger \citep{TheLIGOScientific:2017qsa, GBM:2017lvd, LIGOScientific:2017vwq}. This event established the importance of joint observations of gravitational wave (GW) alerts and identifying electromagnetic (EM) counterparts and investigating the physics of compact object mergers, the neutron star equation of state, and heavy-element production (e.g., \citealt{LIGOScientific:2017vwq, Barnes:2016umi, Coulter:2017wya, Barnes:2013wka, Evans:2017mmy, Goldstein:2017mmi,  Grossman:2013lqa, Haggard:2017qne, Hallinan:2017woc, Kasen:2017sxr, Kasen:2013xka, Kasliwal:2018fwk, Li:1998bw, Margutti:2017cjl, Metzger:2010sy, Metzger:2019zeh,  Roberts:2011xz, Rosswog:2005su, Tanaka:2013ana, Tanvir:2017, Troja:2017nqp}).

The discovery of GW170817 during the second Advanced LIGO–Virgo observing run (O2) marked a breakthrough in multi-messenger astronomy, providing the first confirmed detection of a kilonova powered by r-process nucleosynthesis in the neutron-rich ejecta of a binary neutron star (BNS) merger \citep{TheLIGOScientific:2017qsa, GBM:2017lvd, LIGOScientific:2017vwq}. This event demonstrated the power of coordinated gravitational wave (GW) and electromagnetic (EM) observations for understanding the astrophysics of compact object mergers.

Subsequent analyses of GW170817 placed constraints on the neutron star equation of state \citep[e.g.,][]{Margalit:2017dij, Coughlin:2018miv, Breschi:2021tbm}, confirmed the origin of short gamma-ray bursts \citep[e.g.,][]{Goldstein:2017mmi, Troja:2017nqp}, and provided compelling evidence that neutron star mergers contribute significantly to heavy-element production \citep[e.g.,][]{Li:1998bw, Roberts:2011xz, Rosswog:2005su, Metzger:2010sy, Barnes:2013wka, Kasen:2013xka, Grossman:2013lqa, Tanaka:2013ana, Barnes:2016umi, Kasen:2017sxr, Metzger:2019zeh}. Observational efforts during and after the event provided unprecedented EM coverage across the spectrum \citep[e.g.,][]{Coulter:2017wya, Evans:2017mmy, Haggard:2017qne, Hallinan:2017woc, Margutti:2017cjl, Kasliwal:2018fwk, Tanvir:2017}.

While dozens of GW candidate events with a neutron star have been detected since GW170817, few have been both promising in their classification and well localized enough to enable meaningful EM follow-up. S250206dm, the focus of this study, is notable as the only neutron star merger candidate in the fourth observing run (O4) to date with a 90$\%$ localization area smaller than 600 deg$^2$ and a false alarm rate below one per year \citep{2025GCN.39175....1L, Pillas:2025}. With a $\textgreater 99\%$ probability that one component is a neutron star and a relatively compact skymap released within 36 hours, S250206dm presented one of the most compelling targets for infrared follow-up during O4.

Despite extensive optical and infrared follow-up campaigns, LIGO-Virgo observing runs throughout O3 and O4a yielded no confirmed EM counterparts to BNS or neutron star-black hole (NSBH) mergers \citep[e.g.,][]{Coughlin:2020fwx, Kasliwal:2020, Abbott:2020niy, LIGOScientific:2021qlt, LIGOScientific:2021usb,  Zhu:2021ysz, Pillas:2025}. Many of these events were predicted to be too faint for detection by optical telescopes, as kilonovae exhibit viewing-angle-dependent blue emission that fades rapidly ($<$1 week) \citep{Kasen:2013xka, Kasen:2017sxr, Metzger:2019zeh}. In contrast, the near-infrared (NIR) emission can persist for over a week, with a slower decline timescale and reduced angular dependence due to lanthanide-rich ejecta. This longer duration makes the NIR signal more accessible for follow-up, particularly for events with delayed or large localization. GW170817, for example, had significantly longer-lived infrared emission compared to optical emission \citep{Tanvir:2017}. Models such as \cite{Zhu:2020ffa} predicts kilonovae may be up to $\sim$8–10 times more detectable in the near-infrared compared to optical wavelengths. 

However, systematic sensitive wide-field infrared searches have been limited by high detector costs and high sky backgrounds. Some shallow searches were performed by the Palomar Gattini-IR survey \citep{Moore2016_Gattini, De2020} during O3 \citep{2019ApJ...885L..19C}. Two new wide-field IR surveyors have come online in the fourth observing run, including the Wide-field Infrared Transient Explorer (WINTER), discussed in this study, and PRIME a new NIR survey on a 1.8-meter telescope with four large-format HgCdTe sensors covering a 1.56\,deg$^{2}$ field of view \citep{PRIME_performance, PRIME_camera, PRIME_optics}. WINTER and PRIME, located in the northern and southern hemispheres respectively, offer complementary sky coverage and observing strategies. For example, both followed up the O4a NSBH candidate S240422ed, each covering a substantial portion of the localization region with only partial overlap near its center\footnote{https://treasuremap.space/alerts?graceids=S240422ed}.

WINTER is a 1-meter telescope and camera at Palomar Observatory designed for rapid-response NIR transient discovery, with a wide field of view ($1.2~\text{deg}^2$) and infrared sensitivity ($J_{\rm AB} \sim 18.5$ mag) \citep{Lourie:2020}. WINTER’s InGaAs-based detectors provide a cost-effective alternative to traditional HgCdTe NIR sensors, enabling systematic follow-ua of GW events in $Y$, $J$, and shortened-$H$ bands (0.9-1.7 microns) \citep{Simcoe:2019aps}. %As part of an ongoing effort to expand multi-messenger capabilities in the infrared, WINTER aims to improve the detection prospects for kilonovae missed in previous searches.

WINTER is designed around the discovery of EM counterparts to GW events, with a wide field of view, infrared sensitivity, and rapid robotic follow-up to enable efficient tiling of GW localization regions \citep{Frostig:2020}. \cite{Frostig:2022} presents a comprehensive simulation of WINTER’s capability to follow up BNS mergers during the fourth International Gravitational Wave Network (IGWN) observing run (O4). The analysis predicts that near-infrared kilonovae will persist longer than their optical counterparts, with red kilonovae detectable at distances approximately 1.5 times greater in the NIR. This study offers detailed projections of WINTER’s effectiveness in identifying new kilonovae. However, issues manufacturing the new-to-astronomy InGaAs detectors resulted in a decreased final sensitivity ($J_{\rm AB} \sim 18.5$ instead of $J_{\rm AB} \sim 21.0$ magnitudes), impacting WINTER's effectiveness following up kilonovae \citep{Frostig:2024}. 

%In this paper, we present results from WINTER follow-up of the LIGO--Virgo--KAGRA (LVK; \citealt{TheLIGOScientific:2014jea,TheVirgo:2014hva, Somiya:2011np, Aso:2013eba, akutsu2021}) gravitational wave trigger S250206dm \citep{2025GCN.39175....1L}. S250206dm was first detected by the LVK at 2025-02-06T21:25:30.439 UT, as part of the ongoing O4 run. The event was recovered by the GW search pipelines \texttt{GstLAL} \citep{Messick:2016aqy, Sachdev:2019vvd, Hanna:2019ezx, Cannon:2020qnf, ewing2023performance, Tsukada:2023edh}, \texttt{MBTA} \citep{Adams:2015ulm, Aubin:2020goo} and \texttt{PyCBC Live} \citep{Allen:2005fk, Allen:2004gu, DalCanton:2020vpm, Usman:2015kfa, Nitz:2017svb, Davies:2020tsx}, with an estimated distance of 373\,Mpc and a false alarm rate of about 1 in 25 years. The event, as reported by the LVK, has a 55\% probability of being a NSBH merger, with a 37\% probability of being a BNS merger \citep{2025GCN.39175....1L}; in addition, the probability that the lighter compact object is a neutron star is $>99\%$ \citep{2025GCN.39231....1L}. S250206dm's localization underwent several revisions, with the latest (released 36 hours after merger) having a 50\% localization area of 38\,deg$^2$, and a 90\% area of 547\,deg$^2$. With its relatively precise localization and confident detection, S250206dm is one of the most promising merger candidates in O4 with a high probability of containing a neutron star.

This study presents results from WINTER’s follow-up of the gravitational-wave event S250206dm, detected by the LIGO–Virgo–KAGRA (LVK) collaboration \citep{TheLIGOScientific:2014jea, TheVirgo:2014hva, Somiya:2011np, Aso:2013eba, akutsu2021}. The trigger occurred on 2025 February 6 at 21:25:30.439 UT and was publicly reported via the Gamma-ray Coordinates Network (GCN) \citep{2025GCN.39175....1L}.

The event was identified independently by three low-latency GW search pipelines: \texttt{GstLAL} \citep{Messick:2016aqy, Sachdev:2019vvd, Hanna:2019ezx, Cannon:2020qnf, ewing2023performance, Tsukada:2023edh}, \texttt{MBTA} \citep{Adams:2015ulm, Aubin:2020goo}, and \texttt{PyCBC Live} \citep{Allen:2004gu, Allen:2005fk, Usman:2015kfa, Nitz:2017svb, DalCanton:2020vpm, Davies:2020tsx}. Based on these analyses, S250206dm had an estimated luminosity distance of 373 Mpc and a false alarm rate of roughly 1 in 25 years. According to the LVK’s initial classification, the event has a 55\% probability of being a NSBH merger and a 37\% probability of being a BNS  merger \citep{2025GCN.39175....1L}. Notably, the posterior probability that the lighter compact object is a neutron star exceeds 99\% \citep{2025GCN.39231....1L}. S250206dm’s sky localization was refined over time, with the final skymap—released 36 hours after the merger—yielding a 50\% credible region of 38 deg$^2$ and a 90\% region of 547 deg$^2$. This combination of confident astrophysical classification and relatively compact localization makes S250206dm one of the most promising neutron star merger candidates of the O4 observing run.

In addition to the gravitational wave properties of S250206dm which suggested a possible electromagnetic counterpart, the IceCube Neutrino Observatory \citep{ic_detector} promptly reported the detection of two neutrinos in spatial and temporal coincidence with the event \citep{ic_gcn}, as part of its realtime search for neutrino-gravitational wave coincident detections \citep[see e.g][]{ic_gw}. Neutrino 1 was detected 109s before merger, while neutrino B was detected 289s after merger.\footnote{\url{https://roc.icecube.wisc.edu/public/LvkNuTrackSearch/output/2025_02_06_S250206dm-7-Update.html}} We performed additional ToO observations of neutrino fields, similar to other teams \citep{Becerra2025GCN, Masi2025GCN}. 

In this paper, we report on the non-detection of an EM counterpart to S250206dm, observed with WINTER during O4. We describe our observational strategy, data processing, and implications for future NIR follow-up of GW events.

\section{WINTER follow-up campaign} \label{sec:obs}
\begin{figure*}
    \centering
    \includegraphics[width=0.9\textwidth]{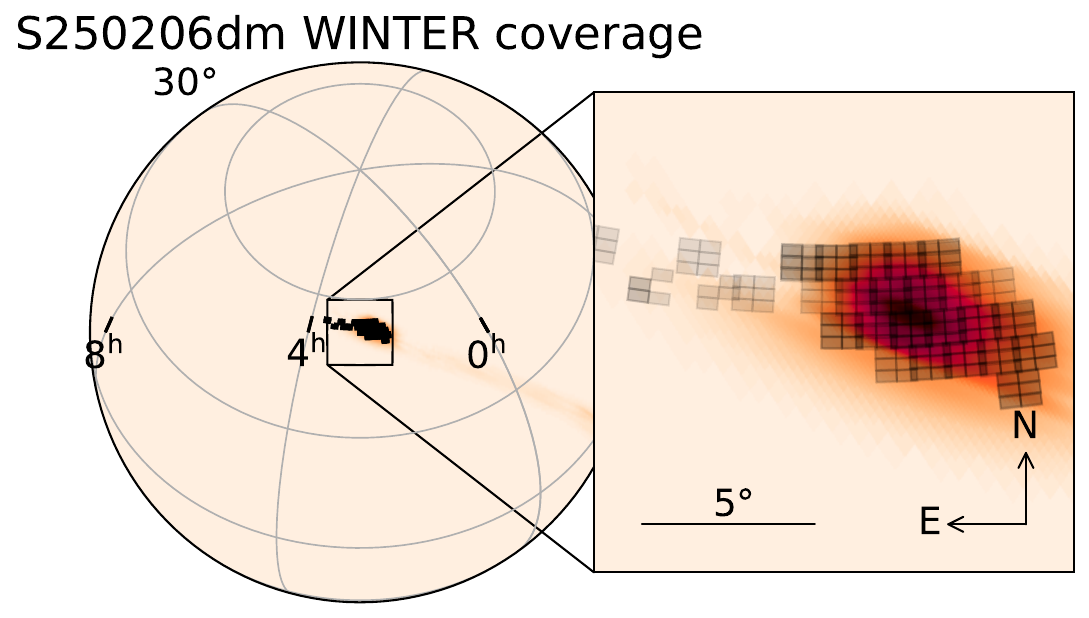}
    \caption{WINTER \emph{J-}band coverage of S250206dm. The probability skymap is plotted in red overlaid with WINTER pointings shown as black rectangles. Each rectangle corresponds to a single WINTER detector and six detectors combine to produce a 1 deg x 1.2 deg FOV in a single WINTER pointing. Our observations covered a total of 43\% of the skymap probability for this gravitational wave event, of which 40\% was observed at least twice. Our observations comprised a total of 39 WINTER fields, a total on-sky time of 130 hours, and achieved median \emph{J-}band limiting magnitudes of 17.6, 17.6, 17.4, 18.0, 17.5, and 17.4 mag (AB) on the six WINTER detectors respectively.}
    \label{fig:coverage_tiling_map}
\end{figure*}

\subsection{WINTER observatory}
The WINTER camera is mounted on a dedicated 1\,m telescope at Palomar Observatory, and comprises six InGaAs detectors that together produce a field-of-view of 1.2\,deg. x 1\,deg with a pixel scale of 1.13\arcsec per pixel. The WINTER telescope has been running robotically on sky since June 2023 \citep{Frostig:2024}, featuring new InGaAs sensor technology and a custom read out \citep{Frostig:2022SPIE, Malonis:2020}, a novel fly's-eye optical design \citep{Lourie:2020}, and custom opto-mechanics \citep{Hinrichsen:2020}. 

WINTER is designed for autonomous, low-latency response to transient alerts, with the ability to begin observations within $\sim$10 minutes of trigger receipt. The telescope is operated by the WINTER Supervisory Program (WSP), a multithreaded \texttt{python}-based control system that handles nightly scheduling, dome operations, and real-time prioritization of Target of Opportunity (ToO) events. ToO requests are submitted through a dedicated API and processed automatically without the need for human intervention.

\subsection{S250206dm observations}
WINTER observations of the GW localization region commenced on UT 2025-02-08T04:05:52, $\approx$30.5 hours after the merger. No observations were possible on the first night after the merger due to poor weather at Palomar Observatory, which continued to limit observations throughout the follow-up campaign. WINTER observations are conducted on a predefined field grid. An optimal schedule to tile the GW skymap using WINTER fields was computed using \texttt{gwemopt} \citep{Coughlin:2018miv, Coughlin:2018fis, Dietrich:2020efo}. Each WINTER field was observed in the \emph{J}-band with a set of eight two-minute long dithered exposures, corresponding to a combined integration time of sixteen minutes per field. These fields were subsequently reobserved on the nights of UT 2025 Feb 9, 10, 11, 16, and 17. All WINTER pointings overlaid on the GW-skymap are shown in Figure \ref{fig:coverage_tiling_map}. The full details of the pointings are available on TreasureMap \citep{Wyatt2020}. Within ten days of merger, a total of 43\% of the probability enclosed within the latest available GW skymap at the time of writing\footnote{\url{https://gracedb.ligo.org/api/superevents/S250206dm/files/Bilby.offline1.multiorder.fits}}  was observed with WINTER at least once, 40\% of the total probability was observed at least twice, and 35\% was observed at least three times.

\subsection{Data processing and transient detection}
The WINTER images were processed by the WINTER data reduction pipeline implemented in \texttt{mirar}\footnote{\url{https://github.com/winter-telescope/mirar}} (Karambelkar \& Stein et al., in prep) -- a modular, \texttt{python}-based framework designed for real-time, end-to-end processing of astronomical images to search for transients. A detailed description of the processing steps will be provided in a forthcoming publication\footnote{See \url{https://mirar.readthedocs.io/en/latest/autogen/winter.default.html} for a flowchart of all steps involved.}. Briefly, the pipeline begins by performing dark subtraction, flat calibration and sky-subtraction on the raw WINTER images. Astrometric solutions are then computed using the packages \texttt{Astrometry.net} \citep{anet} and \texttt{Scamp} \citep{scamp}. Eight astrometry-calibrated dithers are then stacked using the software \texttt{Swarp} \citep{swarp}, and the stacked images are photometrically calibrated relative to the Two Micron All-Sky Survey \citep[2MASS;][]{2mass} point source catalog \citep{Cutri2003}. Images from each of WINTER's six detectors are reduced separately, so each sixteen-minute observation of a field produces six stacked images each with dimensions 0.6\,deg x 0.3\,deg. The median depths achieved during our observations of S250602dm for the six detectors were 17.6, 17.6, 17.4, 18.0, 17.5, and 17.4 mag (AB) respectively. Of the total GW skymap probability, 41\% was observed to a depth of 17 mag or deeper, 38\% was observed to a depth of 17.5 mag or deeper, 31\% was observed to a depth of 18 mag or deeper, and 11\% was observed to a depth of 18.5 mag or deeper. % % The median astrometric uncertainty across all detectors was xx arcsec.

Image subtraction is then performed on these stacked images relative to archival \emph{J-}band reference images from the United Kingdom InfraRed Telescope (UKIRT) Hemisphere Survey \citep[UHS;][]{ukirt_hemisphere} using the optimal image-subtraction algorithm ZOGY \citep{zogy}. Sources identified in the difference image are cataloged in a SQL database, cross-matched to external point-source catalogs such as 2MASS, Pan-STARRS \citep[PS1;][]{Chambers:2016jzn}, as well as to transient alerts from ZTF and the Transient Name Server, packaged as alerts in the \texttt{avro} format, and broadcast to the Fritz.science instance of \texttt{Skyportal} for scanning and visual vetting \citep{skyportal}.  

During the ten day observational campaign, the WINTER images were processed every night in real-time and the alerts were examined to identify interesting transients. Separately, at the end of the campaign, we reprocessed the WINTER images by stacking repeat observations of the same field from multiple nights to increase the depth of our search. The median depths for these ``super-stacks" was 17.8, 17.7, 17.7, 18.0, 17.8, and 17.5 mag for the six detectors respectively.

% \textcolor{red}{This has not yet happened, take out this paragraph?} Beyond this routine processing, we performed additional analysis to identify candidates in areas of sky without reference images. Our observation primarily targeted the northern hemisphere, but some additional observations were accessible to us in the Southern hemisphere. In particular, this southern region included our targeted observations of neutrino B. However, no reference images were available for this southern lobe. Instead, once our primary observing campaign ended, we took deep imaging of these southern fields \textcolor{red}{N} days after merger. We then performed image subtraction using these deep WINTER images as a reference, and performed a retroactive search for candidates. 

\subsection{Candidate searches}

A total of 62991 individual alerts were generated by our standard observations, of which 62675 lay within the 95\% contour reported by LVK. We perform three independent and complementary searches for possible transients in our data and reduce the number of false positive results.

\subsubsection{Candidates with multiple WINTER detections}

Our data reduction pipeline cross-matches each new detection to a dynamically-updated table of WINTER `sources', with each source having an assigned name. The new detections are then crossmatched to all previous detections of the same source, with each \texttt{avro} alert including a complete history of detections in our data. The 62675 alerts were grouped into 57670 unique sources. We select those sources with multiple detections, in order to reject moving objects and reduce the number of bogus sources. Of 57670 sources, we select the 4141 sources with at least 2 detections.  

\subsubsection{Candidates with a ZTF crossmatch}
Every WINTER detection is crossmatched to all archival ZTF alerts \citep{Bellm_2018}, providing a ZTF source name if available. 9608 of the WINTER sources have a pre-assigned ZTF name. For this event, we perform an additional crossmatch on all ZTF alerts detected in the same 2 week period post-merger, providing additional matches in cases where the WINTER detection precede ZTF detections. A total of 116161 ZTF alerts were detected within 14 days of merger in the 95\% contour, for which no additional cuts were applied. These including both ToO and serendiptious detections of the ZTF public survey. The additional crossmatch yields 1275 matches, of which 65 are new. In total, 9673/57670 WINTER sources have a ZTF crossmatch. 

% We only require a single WINTER detection, with the time separation to coincident ZTF detection being sufficient to reject moving object.  X\% of the WINTER .  Directly cross-matching these two source lists with a radius of 1.5" yields N sources with both ZTF+WINTER detections.

\begin{table*}[]
    \centering
    \begin{tabular}{c|c|c|c|c}
        Name & RA& Dec & Date & Limiting \\
         & &  &  &  Magnitude [AB]\\
        \hline
        % AT 2025bmq & &&&&& Nuclear transient. \\
        % & &&&&& Nuclear transient. \\
        \hline
        AT2025bmq & 38.518 & 54.572 & 2025-02-08T06:44:50.635 & 18.3 \\
         &  &  & 2025-02-10T05:15:34.858 & 17.7  \\
        \hline
        AT2025bbp & 153.782 & -22.882 & 2025-02-08T11:04:14.707 & 17.3 \\
         &  &  & 2025-02-11T09:29:20.609  & 17.0\\
        \hline
        AT2025ban & 152.918 & -18.952 & 2025-02-09T08:58:45.274 & 14.4 \\
        \hline
        AT2025baj & 156.517 & -28.346 & 2025-02-08T11:21:47.004 & 14.6 \\
         &  &  & 2025-02-11T10:39:07.412 & 17.3 \\
    \end{tabular}
    \caption{Summary of WINTER images which overlap transients reported to TNS. The image-wide limiting magnitude is also reported, though all transients had underlying hosts which complicate recovery.}
    \label{tab:tns_table}
\end{table*}
\subsubsection{Candidates with a TNS cross-match}
We also check against the Transient Name Server, which serves as the official community repository of all astrophysical transients.
We select all candidates reported to TNS which lie within the S250206dm skymap (accessed via the dedicated TNS event page for S250206dm). There were 116 at the time of writing.\footnote{\url{https://www.wis-tns.org/system/files/ligo/o4/S250206dm_20250206_212530/S250206dm_20250206_212530_AFTER.json}}. Of these, 95 lie within the 95\% contour, though just four TNS transients fell within fields observed by WINTER. We crossmatch all WINTER candidates to the TNS transients with a radius of 3", but find no coincident detections. 

In Table \ref{tab:tns_table} we list each WINTER image which overlapped with a TNS transient. Of the four TNS transients, only AT2025bmq fell in a region observed by WINTER and had $J$-band reference images, both necessary requirements to generate an alert. As noted in the table, no such alert was produced. AT2025bmq is coincident with an IR-bright host which is clearly recovered in the WINTER images at $J_{\rm AB} \sim 14.5$. This is consistent with the archival magnitude of this source, so we cannot resolve whether there is any excess flux from a nuclear transient.

For the other three sources (AT2025bbp, AT2025ban and AT2025baj), no reference images were available, so no image subtraction was performed. All three have apparent underlying hosts, but given the depths of our images, neither the host nor transient were recovered in our science images. Only one of these three transients (AT2025baj) has a host bright enough to be detected in 2MASS. However,  at $J_{AB}$=17.4 mag, this host is fainter than the limiting magnitudes of our science images. 

\begin{table*}[]
    \centering
    \begin{tabular}{c || c|c|c||c}
         & WINTER + WINTER & WINTER + ZTF & WINTER + TNS & Combined Total\\
         \hline
         Initial Sources &4141 &9673& 0& 11802\\
         After Stellar Cuts & 140 &140& 0&274\\
         After WINTER predection cuts &140&140&0&274\\
         After ZTF predection cuts &135&4&0&139\\
         \hline
         After human vetting &0&0&0&0\\
    \end{tabular}
    \caption{Statistics for source cuts using the three independent candidate search methods. Given that sources can be selected by multiple searches, we also give the combined total in the right-most column. We reject sources which have either a WINTER or ZTF detection preceding the merger.}
    \label{tab:Cuts Summary}
\end{table*}

\subsubsection{Candidate Vetting}

Though many sources remain at this stage, the vast majority will be stellar. For all 11802 candidates selected by our three independent methods, we perform a series of algorithmic cuts:

\begin{itemize}
    \item We crossmatch all sources to PS1 to a star/galaxy morphology classifier trained using PS1 data \citep{sgscore}. We reject WINTER sources within 7" of bright (m$_{r}<15$) stars, and those within 3" of fainter stars. 
    \item We reject sources within 3" of sources detected in Gaia-DR3 \citep{gaia_dr3} with $>3\sigma$ parallax. 
    \item We finally remove sources within 20" of very bright Gaia-DR3 stars, defined as M$_{G}<14$.
    \item We reject sources which are `old', defined as having either a WINTER or ZTF detection prior to merger.
\end{itemize}

For sources passing these cuts, we perform visual inspection. We were aided  by crossmatches to Gaia-DR3 \citep{gaia_dr3}, SDSS \citep{sdss_00} and WISE \citep{wise} for identifying known variable sources. We reject candidates based on the following criteria:

\begin{itemize}
    % \item Candidates with significant ($>3 \sigma$) measured parallax in Gaia-DR3 \citep{gaia_dr3} are classified as stellar
    \item Candidates which are nuclear are classified as probable AGN if the host galaxy is either a spectroscopic AGN in SDSS \citep{sdss_00}, is listed as a quasar in Milliquas \citep{milliquas} or has AGN-like WISE colours of W1-W2>0.7 \citep{stern_12}.
    \item Bogus sources with poor subtraction residuals, especially those detections close to poorly-subtracted reference image sources. We require candidates to have at least two detections which pass visual inspection and appear to be plausibly real. A handful of candidates which nominally had two detections were rejected at this stage after careful visual comparison to UKIRT reference images and other sources within the field, on the grounds that at least one of the apparent detections was spurious.
\end{itemize}

On this basis, every WINTER candidate is rejected as a possible transient. A full summary of cut statistics is given in Table \ref{tab:Cuts Summary}.

\section{Comparison with kilonova models}

%\begin{figure*}[]
%    \centering
%    \includegraphics[width=0.5\textwidth]{figures/100distances.pdf}
%    \caption{NSBH kilonova light curve models from \cite{Bulla:2024}, marginalized over the full parameter grid and 100 distances sampled from the event probability distribution function for the estimated distance. The median light curve across all distance realizations and the full model grid is shown as a red line, while the shaded regions represent the 1$\sigma$ (68$\%$) and 3$\sigma$ (99.7$\%$) credible intervals. The limiting magnitudes of WINTER observations are shown for comparison and do not constrain any of the presented models.   }
%    \label{fig:model_lc}
%\end{figure*}

We generate model light curves for kilonovae using NSBH and BNS models presented in Ahumada et al., in prep. These models are simulated with the latest version \citep{Bulla:2023} of the 3D Monte Carlo radiative transfer code \textsc{possis} \citep{Bulla:2019muo} assuming a two-component (dynamical+wind) ejecta configuration and using state-of-the-art heating rates \citep{Rosswog:2024} and opacities \citep{Tanaka:2020}. The kilonova grids span the expected parameter space, with the NSBH grid varying the average velocity of the dynamical ejecta and the masses of the dynamical and wind ejecta (for a total of 108 models) and the BNS grid the mass, average velocity and average electron fraction for both components (for a total of 3072 models). To account for distance uncertainties, we sample 100 distances from the probability distribution function for S250206dm, as provided by the multi-order \texttt{Bilby} skymap \citep{2025GCN.39231....1L}, resulting in a distribution with a mean of 375 Mpc and a 1$\sigma$ (68$\%$) width of 107 Mpc. For each sampled distance, we compute a full model grid of calculated ejecta properties to explore the range of possible emission scenarios.

\begin{figure*}[]
    \centering
    \begin{minipage}{0.48\textwidth} 
        \centering
        \includegraphics[width=1.1\linewidth]{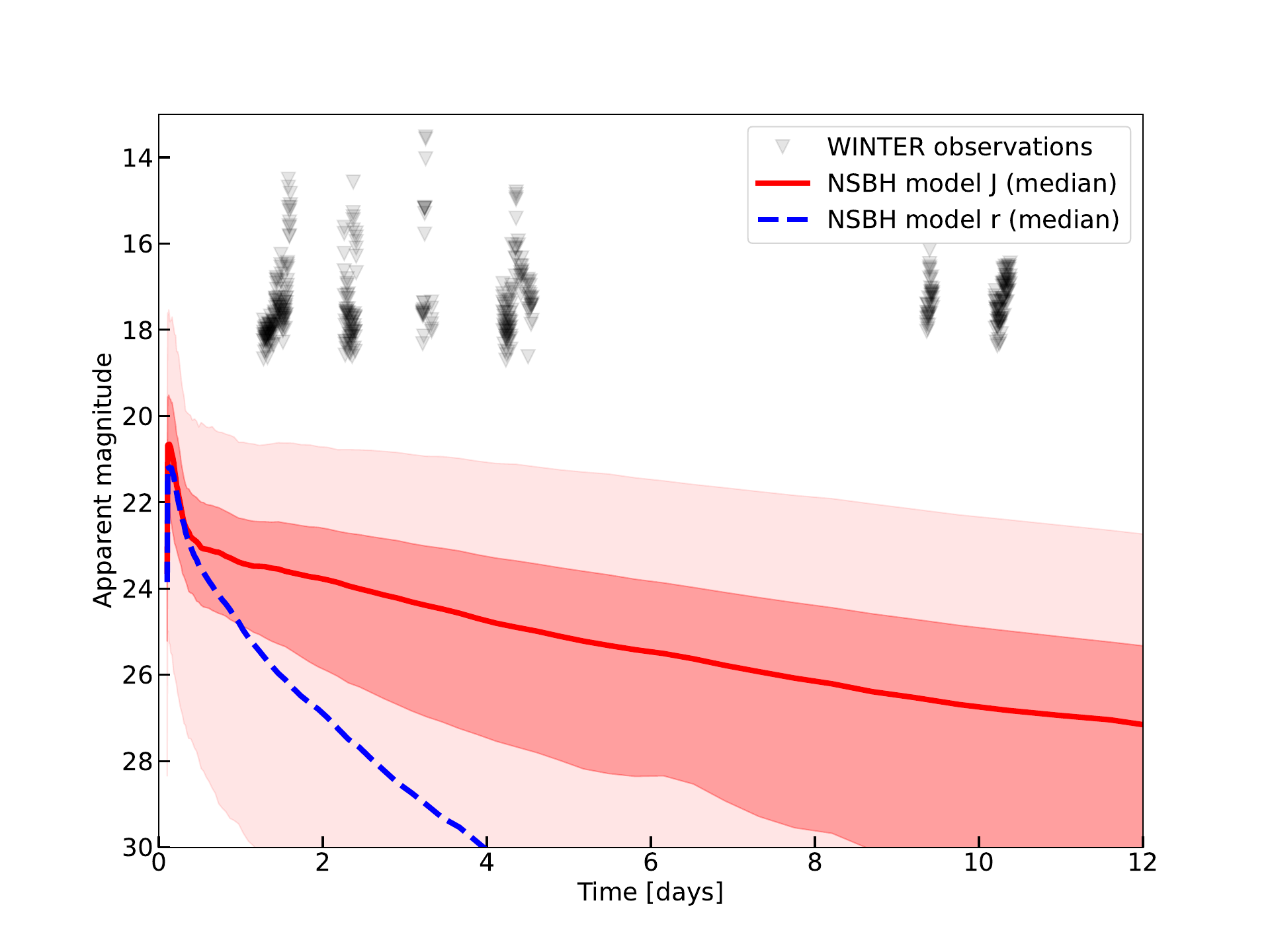}
    \end{minipage}
    \hspace{0mm} 
    \begin{minipage}{0.48\textwidth}
        \centering
        \includegraphics[width=1.1\linewidth]{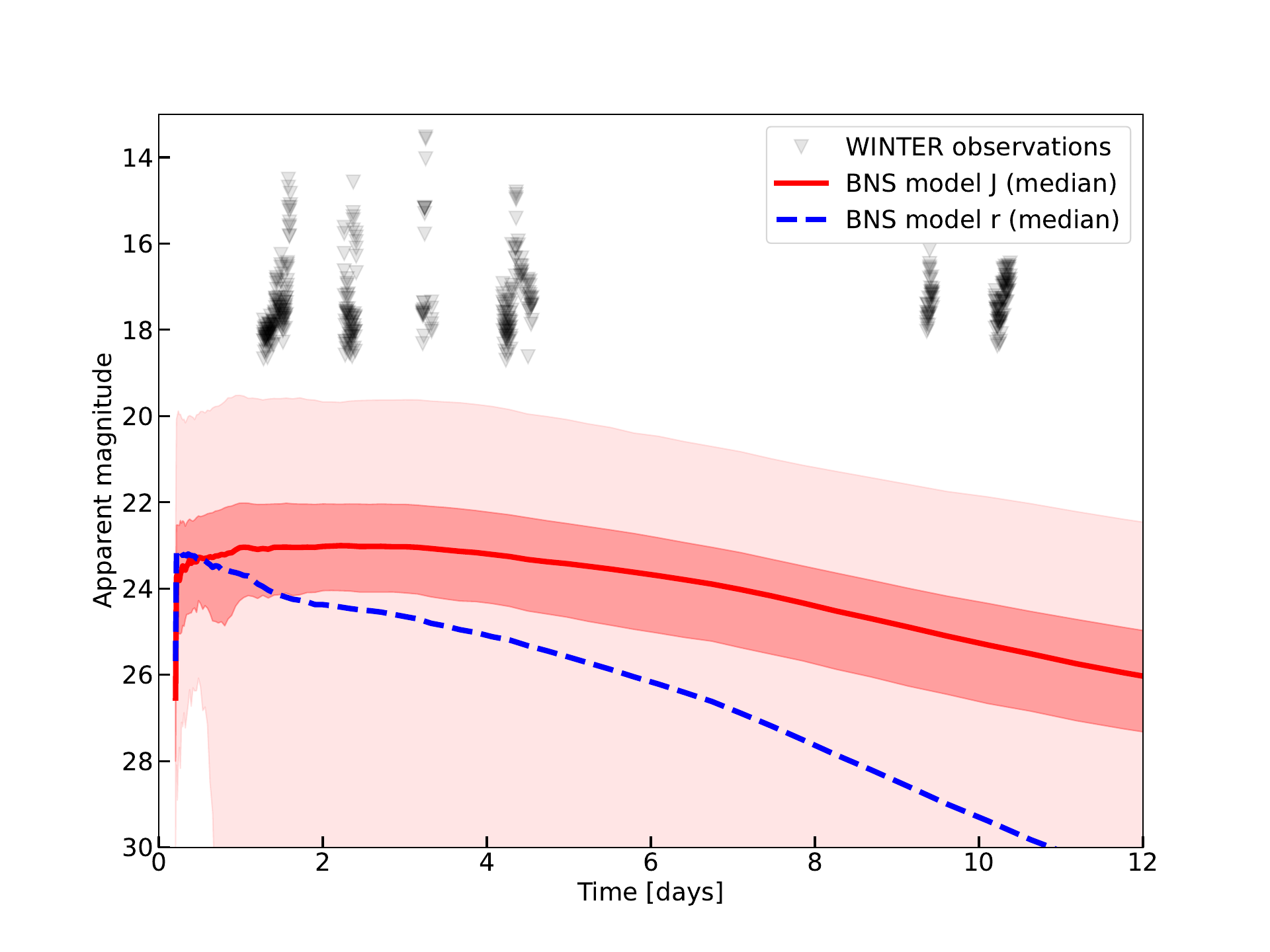}
    \end{minipage}
    \caption{WINTER observations compared to kilonova models for NSBH (left) and BNS (right) mergers from Ahumada et al., in prep, computed using the radiative transfer code \textsc{possis} \citep{Bulla:2023}. The kilonova light curve models are marginalized over the full parameter grid and 100 distances sampled from the event probability distribution function for the estimated distance. The median $J$-band light curve across all distance realizations and the full model grid is shown as a red line, while the shaded regions represent the 1$\sigma$ (68$\%$) and 3$\sigma$ (99.7$\%$) credible intervals. For comparison, the $r$-band light curve is shown from the same exercise to demonstrate the extended NIR plateau in kilonova models. The limiting magnitudes of WINTER observations are shown for comparison and do not constrain any of the presented models. }
    \label{fig:model_lc}
\end{figure*}

Figure \ref{fig:model_lc} presents the combined model prediction, obtained by interpolating all light curves onto a common time grid and computing summary statistics at each epoch. We present the $J$-band median light curve and confidence intervals and compare to the $r$-band median light curve to demonstrate the relatively bright and persistent NIR plateau. Additionally, we overlay the limiting magnitudes of all WINTER follow-up observations (as described in Section \ref{sec:obs}) to assess whether the models are observationally constrained. 

The absence of a detected kilonova in WINTER observations is consistent with the predicted NSBH and BNS model grids. At the estimated distance of S250206dm and the timing of the observations, the WINTER observations are not sufficiently deep to place meaningful constraints on any of the kilonova models explored in this study.

WINTER's effectiveness in gravitational-wave counterpart searches depends on both weather conditions and achieved sensitivity. The follow-up campaign for S250206dm was hindered by poor weather, which prevented early observations on the first night after the GW trigger. The delay of 30.5 hours resulted in a decline of 3.1 and 2.9 magnitudes from peak brightness for the NSBH and BNS models, respectively. Under clear conditions, WINTER can respond robotically to a trigger in under ten minutes. Had WINTER been able to observe S250206dm within the first hour, it could have constrained some of the NSBH models presented here within the 3$\sigma$ credible interval, though it would still lack the depth to constrain any BNS models or a typical NSBH merger at 373 Mpc. In clear observing conditions, WINTER can achieve stacked $J$-band depths of 19.5 mag, compared to only 18.0 mag in this weather-limited campaign. This difference significantly impacts the search horizon: the peak of the median NSBH (BNS) model shown would be detectable out to 219 (74) Mpc, compared to just 109 (37) Mpc under poorer conditions. Furthermore, issues with manufacturing the WINTER sensors lead to a significant decrease in achievable depth from the design (see \cite{Frostig:2024} for more details). At the designed limiting magnitude of $J_{\rm AB}$ = 21  mag, WINTER would be able to constrain the peak of the median NSBH (BNS) merger model to 437 (149) Mpc.

WINTER's non-detection of 250206dm is consistent with the non-detections reported throughout O3 and O4a. During O3, most BNS and NSBH merger candidates were localized to large areas (median 4480 $\text{deg}^2$) and occurred at significant distances (median 267 Mpc), with many events $\textgreater$300 Mpc \citep{Kasliwal:2020} Despite rapid optical follow-up, no kilonova counterparts were found. Notably, WINTER was not yet operational during O3, and no wide-field facility conducted systematic NIR observations to comparable depths. The lack of NIR sensitivity would have limited detectability for red, lanthanide-rich kilonovae or those viewed at higher inclinations. Similarly, during O4a, 13 BNS and NSBH mergers yielded no counterparts despite extensive optical and infrared coverage \citep[e.g.,][]{Ahumada:2024, Pillas:2025}.

This null result does not constrain kilonova model parameters, such as ejecta mass or inclination. However, future events with more favorable observations will allow joint optical and NIR campaigns to rule out portions of the ejecta parameter space. \citet{Pillas:2025} presents a framework for such analyses, combining multi-wavelength non-detections to constrain ejecta mass and viewing angle in well-localized mergers. Although the WINTER observations used in that study of O4 NSBH candidate S240422ed were too shallow to contribute meaningful constraints, the study highlights the value of coordinated, multi-band follow-up.

\section{Summary and way forward}
In this study, we present results from WINTER's NIR follow-up campaign of the gravitational-wave event S250206dm, a candidate NSBH or BNS merger detected during the fourth LIGO-Virgo-KAGRA GW observing run. We conducted a targeted follow-up campaign to search for a potential EM counterpart. With systematic coverage of $43\%$ of the probability region within ten days post-merger, we recover no EM counterpart. 

The non-detection is consistent with the modeled light curves of kilonovae and our observational depths (median $J_{\rm AB} \sim 17.4 - 18.0$ mag by WINTER sensor), which were insufficient to place strong constraints on kilonova models at these distances. However, this campaign demonstrates that under improved conditions—e.g., median $J_{\rm AB} \gtrsim 19.5$ mag and response within 1–2 hours post-merger—WINTER could constrain or rule out the brighter end of predicted NSBH kilonova models, particularly those with high ejecta masses and polar viewing angles. For example, WINTER can rule out red kilonovae similar to GW170817 for well-localized events within $\sim$200 Mpc, enabling constraints on ejecta mass, composition, and geometry for a meaningful fraction of future NSBH or BNS detections. %Furthermore, no confirmed counterpart was reported by other optical or infrared facilities, highlighting the challenges of detecting EM counterparts to distant compact object mergers with current facilities.

WINTER's null result is consistent with non-detections reported by other optical and infrared facilities for this event. No confirmed EM counterpart to S250206dm was identified, underscoring the challenges of detecting kilonovae from distant mergers with current instruments and the importance of coordinated deep optical and NIR campaigns.

This study highlights WINTER’s unique value as the first dedicated wide-field NIR transient survey optimized for rapid gravitational-wave follow-up. Efficient coverage of the gravitational-wave sky localization areas remains a critical challenge for multi-messenger follow-up. WINTER's wide field of view (1 $\text{deg}^2)$ and rapid robotic response ($\sim$10 minutes) help quickly tile the localization area. Furthermore, WINTER adds deep NIR coverage ($J_{\text{AB}}$>18.5 mag), which was not available during previous gravitational-wave runs. This enables sensitivity to kilonova emission scenarios with redder, lanthanide-rich ejecta and extends the effective search window by probing the longer-lived NIR plateau characteristic of many kilonova light curves. These capabilities fill a critical gap in multi-messenger follow-up, which was unaddressed during O3 and remains underdeveloped even in O4a.

The development of future instruments with greater sensitivity, wider fields of view, and improved sky coverage will be critical for detecting and characterizing kilonovae, especially as the sensitivity of gravitational-wave detectors improve. The upcoming Cryoscope surveyor in Antarctica will cover a 50 $\text{deg}^2$ field of view in the $K$-dark to a depth of 24.1\,mag AB in one hour \citep{Cryoscope} and the Nancy Grace Roman Space Telescope will provide a depth of 27.60 mag in an hour in the F129 filter over a field of 0.281~$\text{deg}^2$.\footnote{\url{https://roman.gsfc.nasa.gov/science/WFI_technical.html}} Furthermore, more precise and rapid gravitational wave sky localizations from future observing runs will improve targeting strategies for electromagnetic counterpart searches. Finally, next-generation ground-based observatories, such as the Vera C. Rubin Observatory \citep{Rubin}, will offer complementary optical coverage, further expanding the capability to identify and study transient astrophysical phenomena.

Although WINTER did not detect a counterpart to S250206dm, this follow-up campaign demonstrates a systematic NIR multi-messenger approach. Future attempts, coupled with advancing technology, will further improve our odds of identifying and characterizing the electromagnetic signatures of compact object mergers, deepening our understanding of neutron star physics, nucleosynthesis, and the transient universe.

\section*{Acknowledgments}
WINTER’s construction is made possible by the National Science Foundation under MRI grant number AST-1828470 with early operations supported by AST-1828470. Significant support for WINTER also comes from the California Institute of Technology, the Caltech Optical Observatories, the Bruno Rossi Fund of the MIT Kavli Institute for Astrophysics and Space Research, the David and Lucille Packard Foundation, and the MIT Department of Physics and School of Science. D.F.'s contribution to this material is based upon work supported by the National Science Foundation under Award No. AST-2401779. This research award is partially funded by a generous gift of Charles Simonyi to the NSF Division of Astronomical Sciences. The award is made in recognition of significant contributions to Rubin Observatory’s Legacy Survey of Space and Time. M.B. acknowledges the Department of Physics and Earth Science of the University of Ferrara for the financial support through
the FIRD 2024 grant.

\bibliography{myreferences}
\end{document}